\documentstyle[aps,psfig]{revtex}

\newcommand{\beq}{\begin{equation}}
\newcommand{\eeq}{\end{equation}}

\newcommand{\NN}{{\mathcal{N}}}
\newcommand{\NP}{N_{\cal P}}
\newcommand{\KP}{K_{\cal P}}
\newcommand{\UP}{U_{\cal P}}

\newcommand{\RR}{{\bf R}}
\newcommand{\RRp}{{\bf R'}}

\newcommand{\rr}{{\bf r}}

\newcommand{\mm}{{\bf m}}

\newcommand{\sss}{{\bf s}}
\newcommand{\HH}{{\mathcal{H}}}

\newcommand{\BE}{\begin{equation}}
\newcommand{\EE}{\end{equation}}
\newcommand{\BEN}{\begin{eqnarray}}
\newcommand{\EEN}{\end{eqnarray}}
\newcommand{\erf}{\mbox{erf}}
\newcommand{\erfc}{\mbox{erfc}}
\newcommand{\Tr}{\mbox{Tr}}

\newcommand{\wt}{\tilde{w}}

\begin{document}
\title{Variational Density Matrix 
Method for Warm Condensed Matter and Application to Dense Hydrogen}
 
\author{Burkhard Militzer$^{a)}$  and E. L. Pollock$^{b)}$}
\address{$^{a)}$Department of Physics \\
University of Illinois at Urbana-Champaign, Urbana, Illinois 61801\\
$^{b)}$Physics Department,\\
         Lawrence Livermore National Laboratory,\\
         University of California, Livermore, California 94550 }
 
\date{\today}
\maketitle
 
\begin{abstract}

   A new variational principle for optimizing thermal density matrices is
introduced. As a first application, the variational many body density 
matrix is written as a determinant of one body density matrices, which
are approximated by Gaussians with the mean, width and amplitude as 
variational parameters. The method is illustrated for the particle in an 
external field problem, the hydrogen molecule and dense hydrogen where the
molecular, the dissociated and the plasma regime are described. Structural
and thermodynamic properties (energy, equation of state and shock Hugoniot) are presented.
\end{abstract}

\section{Introduction}

      Considerable effort has been devoted to systems where finite
temperature ions (treated either classically or quantum mechanically
by path integral methods) are coupled to  degenerate electrons on the
Born-Oppenheimer surface. In contrast, the theory for similar systems
with non-degenerate electrons ($T$ a significant fraction of $T_{Fermi}$)
is relatively underdeveloped except at the extreme high $T$ limit
where Thomas-Fermi and similar theories apply. In this paper  we
present a computational approach for systems with non-degenerate
electrons analogous to the methods used for ground state many body
computations.

      Although an oversimplification, we may usefully view the
ground state computations as consisting of three levels of
increasing accuracy \cite{Ha94}. At the first level, the ground state wave function
consists of determinants, for both spin species, of single particle
orbitals often taken from local density functional theory
\beq
\Psi_{GS}(\RR)=\left|
\begin{array}{ccc}
\Phi_{1}(\rr_{1})&\ldots&\Phi_{N}(\rr_{1})\\
\ldots&\ldots&\ldots\\
\Phi_{1}(\rr_{N})&\ldots&\Phi_{N}(\rr_{N})\\
\end{array}\right| \quad.
\eeq
The majority of ground state condensed matter calculations stop at this
level.

If desired, additional correlations may be included by multiplying
the above wave function by a Jastrow factor,
$\prod_{i,j}f(r_{ij})$, where the $f$ will also depend
on the type of pair (electron-electron, electron-ion). Computing
expectations exactly (within statistical uncertainty), with this type of
wave function now requires Monte Carlo methods.

Finally diffusion Monte Carlo \cite{CM96,FM99} methods using the nodes of this wave function to
avoid the Fermion problem may be used to calculate the exact
correlations consistent with the nodal structure.

The finite temperature theory proceeds similarly. Rather than the
ground state wave function a thermal density matrix
\beq
    \label{eq2}
\rho(\RR,\RRp;\beta)= \left< \RR | e^{-\beta \HH  }|\RRp \right> =
\sum_{s}e^{-\beta E_{s}}\Psi_{s}(\RR)\Psi_{s}(\RRp)
\eeq
is needed to compute the thermal averages of operators
\beq
\left< {\bf O}\right> = \frac{ \Tr\left[ {\bf O}\rho \right] }{\Tr\left[ \rho \right]}
\;. 
\label{thermal_average}
\eeq

At the first level, this many body density matrix may be approximated
by determinants of one-body density matrices, for both spin types,
as well as the ions
\beq
\rho(\RR,\RRp;\beta)=\left|
\begin{array}{ccc}
\rho_1(r_{1},r'_{1};\beta)&\ldots&\rho_1(r_{N},r'_{1};\beta)\\
\ldots&\ldots&\ldots\\
\rho_1(r_{1},r'_{N};\beta)&\ldots&\rho_1(r_{N},r'_{N};\beta)
\end{array}\right| 
                                     \label{eq:1.3}
\eeq

The Jastrow factor can be extended to finite temperature and the
above density matrix multiplied by $\prod_{i,j}f(r_{ij},r'_{ij};\beta)$.
In particular, the high temperature density matrix used in path integral
computations has this form. 

Finally, the nodal structure from this variational density matrix (VDM)
may be used in restricted path integral Monte Carlo (RPIMC) \cite{Ce91,PC94,Ce95,Ce96,Ma96}. 
This method has been extensively applied
using the free particle nodes. One aim of the present work is
to provide more realistic nodal structures as input to RPIMC.

   This paper considers the first level in this approach. The next section is devoted
to a general variational principle which will be used to determine the many body density
matrix. The principle is then applied to the problem of a single particle in an external 
potential and compared to exact results for the hydrogen atom density matrix. After a 
discussion of some general properties, many body applications are considered starting
with a hydrogen molecule and then proceeding to warm, dense hydrogen. It is shown that the
method and the ansatz considered can describe dense hydrogen in
the molecular, the dissociated and the plasma regime. Structural and thermodynamic 
properties for this system over a range of temperatures (T$=5\,000$ to $250\,000\,K$) and
densities (electron sphere radius $r_s=1.75$ to $4.0$) are presented.

\section{Variational Principle for the Many Body Density Matrix}
\label{variational}

      The Gibbs-Delbruck variational principle for the
free energy based on a trial density matrix
\beq
F\leq \Tr[\tilde{\rho}\HH ]+kT\;\Tr[\tilde{\rho}\ln\tilde{\rho}]
\eeq
where
\beq
\tilde{\rho}=\rho/\Tr[\rho]
\eeq
is well known and convenient for discrete systems (e.g. Hubbard models) but the logarithmic 
entropy term makes it difficult to apply to  continuous systems. 
Here, we propose a simpler variational
principle patterned after the Dirac-Frenkel-McLachlan variational principle
used in the time dependent quantum  problem \cite{Mc64}. Consider the quantity
\beq
                    \label{gen_var}
I\left(\frac{ \partial \rho}{\partial \beta}\right)=
\Tr\left(\frac{ \partial \rho}{\partial \beta} + \HH \rho\right)^{2}
\eeq
as a functional of
\beq
\Theta\equiv \frac{ \partial \rho}{\partial \beta}
\eeq
\beq
I\left(\Theta\right)=
\Tr\left(\Theta + \HH \rho\right)^{2}
\eeq
with $\rho$ fixed.
$I\left(\Theta\right)=0$ when $\Theta $ satisfies the Bloch equation,
 $\Theta=-\HH\rho$,
and is otherwise positive. Varying $I$ with $\Theta $
gives the minimum condition
\beq
\Tr\;\left[ \delta\Theta\left(\Theta +\HH \rho\right)\right]=0 \quad.
\eeq
This may be written in a real space basis as
\beq
\label{no_symmetry_assumed}
\int\int \delta\Theta(\RRp,\RR;\beta)
\left[\Theta(\RR,\RRp;\beta)+\HH \rho(\RR,\RRp;\beta)
\right] d\RR d\RRp=0
\eeq
or, using the symmetry of the density matrix in $\RR$ and $\RRp$,
\beq
\label{symmetry_assumed}
\int\int \delta\Theta(\RR,\RRp;\beta)
\left[\Theta(\RR,\RRp;\beta)+\HH \rho(\RR,\RRp;\beta)
\right] d\RR d\RRp=0 \quad.
\eeq
Finally, we may consider a variation at some arbitrary, fixed $\RRp$
to get
\beq
\label{symmetry_assumed2}
\int \delta\Theta(\RR,\RRp;\beta)
\left[\Theta(\RR,\RRp;\beta)+\HH \rho(\RR,\RRp;\beta)
\right] d\RR =0\;\;\forall \RRp.   \label{eq:2.11}
\eeq
It should be noted that in going from Eq. \ref{no_symmetry_assumed} to 
Eq. \ref{symmetry_assumed} a density matrix symmetric 
in $\RR$ and $\RRp$ is assumed, which is a property of the exact 
density matrix.
If the variational ansatz does not manifestly have this invariance
Eq. \ref{symmetry_assumed2} minimizes the quantity,
\beq
\int 
\left[\Theta(\RR,\RRp;\beta)+\HH \rho(\RR,\RRp;\beta)
\right]^{2} d\RR=0 \quad.
\eeq
We propose solving this equation by parameterizing
the density matrix with a set of parameters $q_i$ depending on
imaginary time $\beta$ and $\RRp$,
\beq
\rho(\RR,\RRp;\beta)=\rho(\RR,q_1,\ldots,q_m)
\;\;\mbox{where}\;\;q_{i}(\RRp;\beta)
\eeq
so
\beq
\Theta(\RR,\RRp;\beta)=\sum_{i=1}^m 
   \frac{\partial q_i(\RRp;\beta)}{\partial \beta}
   \frac{\partial \rho (\RR,{q})}{\partial q_i}
=\sum_{i=1}^m \dot{q}_i \: \frac{\partial \rho}{\partial q_i}\;.
                     \label{eq:2.13}
\eeq
 In the imaginary time derivative
$\Theta$ only variations in $\dot{q}$ and not ${q}$  are considered since
$\rho$ is fixed so,
\beq
\delta \Theta(\RR,\RRp;\beta) = \sum_{i=1}^m \delta \dot{q}_i(\RRp;\beta)
 \: \frac{\partial \rho(\RR,{q})}{\partial q_i} \quad.
\eeq
Using this in equation~\ref{eq:2.11} gives for each
variational parameter, since these are independent,
\beq
\int \!  \frac{\partial \rho}{\partial q_j} \left(\Theta +  \HH \rho \right)
         d\RR  = 0\;\;. \label{eq:2.15}
\eeq
This reveals the imaginary-time equivalent to the approach of 
Singer and Smith \cite{SS86}
for an approximate solution  of the time dependent Sch\"odinger equation using wave packets 
(see section~\ref{realtime}). Introducing the notation 
\beq
            p_i \equiv \frac{\partial(\mbox{ln} \rho)}{\partial q_i}
                                                           \label{eq:2.16}
\eeq
and using Eq. \ref{eq:2.13},
the fundamental set of first order differential equations
for the dynamics of the variation parameters in imaginary time
follows from Eq.. \ref{eq:2.15} as,

\beq
\label{eq:2.17}
  \int\! p_j \: \rho \HH \rho\;d\RR \:\; +
  \sum_{i=1}^m \dot{q}_i \int\! p_j \: p_i \: \rho^2 \; d\RR 
   \;\; = \;\; 0
\eeq
or in matrix form
\beq
   \label{matrix}
\frac{1}{2}\frac{\partial H}{\partial \vec{q}}\; + \;\, 
    \stackrel{{\textstyle \leftrightarrow}}{\NN}\: \dot{\vec{q}}=  0
\eeq
where
\beq
H \equiv \int \rho \HH \rho\;d\RR  \label{H}
\eeq
and the norm matrix 
\beq
\NN_{ij}\equiv \int p_i \: p_j \: \rho^2 \, d\RR =\lim_{q'\rightarrow q}
\frac{\partial^2 N}{ \partial q_i \partial q'_j}  \label{NN}
\eeq
with
\beq
               \label{norm_gen}
N \equiv \int \rho(\RR,\vec{q}\,;\beta) \; \rho(\RR,\vec{q}\:'\,;\beta)\;d\RR\;\;.
\eeq
The initial conditions follow from the free particle limit of the
density matrix at high temperature, $\beta\rightarrow 0$,
\beq
      \label{fpl}
\rho(\RR,\RRp;\beta)\rightarrow \exp\left[ -(\RR-\RRp)^{2}/4 \lambda \beta\right]
     /(4\pi\lambda\beta)^{3N/2} \quad \mbox{where} \quad \lambda = \hbar^2/2m\quad.
\eeq
Various ansatz forms for $\rho$ may now be used with this approach.
After considering the analogy to real time wave packet molecular dynamics, the
principle is first applied to the problem of a particle in an external field.

\section{Analogy to real-time wave packet molecular dynamics}
\label{realtime}

Wave packet molecular dynamics (WPMD) was first used by Heller\cite{He75} and later 
applied to scattering processes in nuclear physics \cite{F90}
and plasma physics \cite{KTR94b,EM97}.
An ansatz for the wave function $\psi({q_\nu})$ is made and the 
equation of motions 
for the parameters ${q_\nu}$ in real time can be derived from the 
principle of stationary action  \cite{F90},
\beq
\delta \int dt \:
L = 0
\quad,\quad
L \left({q_\nu}(t),{\dot{q}_\nu}(t) \right) =
\left\langle \psi
\left| i {\partial}_t 
- \HH \right| 
\psi \right\rangle
\label{variational_principle}
\eeq
This leads to a set of first order equations, which provides an approximate 
solution of the Schr\"odinger equation.
However, this principle cannot be directly applied to the Bloch equation because there is no
imaginary part in the density matrix. For this reason, we followed in our derivation in
section \ref{variational} the principle of Dirac, Frenkel and McLachlan \cite{Mc64}, 
which minimizes the quantity
\beq
  \int | \HH \psi - i \hbar \theta | ^2 \, dt,\quad\theta = \frac{\partial \psi}{\partial t}\;.
\eeq
This method was employed in 
\cite{SS86} to obtain the dynamical equations in real time.

The VDM approach and WPMD method share the zero temperate limit, 
which is given by the  Rayleigh-Ritz principle 
(see section \ref{ZeroTLimit}).
At high temperature, the width of wave packets in WPMD grows without limits,
which is a known problem of this method \cite{Mi96b,KRT98}. In
the VDM approach, the correct high temperature limit of free particles
is included. The average width shown in Fig. \ref{averageWidth} can be used to 
verify the attempts to correct the dynamics of the real time wave packets in \cite{KRT98}.

\section{Example: Particle in an external field}
As a first example, we apply this method to the problem of one particle
in an external potential
\beq
\HH = -
\lambda
{\bf \nabla}^{2} + V(r)\;\;.
\eeq
 The one-particle density matrix will be 
approximated as a Gaussian with the mean $\mm$, width $w$ and 
amplitude factor $D$,
\beq
                 \label{gauss}
\rho_1(\rr,\rr',\beta) = (\pi w)^{-3/2} \: \mbox{exp} 
\left\{ -\frac{1}{w} (\rr-\mm)^2 + D \right\}
\eeq
as variational parameters.
The initial conditions at $\beta\longrightarrow 0$ are 
$w= 4\lambda\beta$, $\mm=\rr'$ and $D=0$
in order to regain the correct free particle limit, Eq. \ref{fpl}.
For this ansatz $H$, defined in Eq. \ref{H} as
\beq
H\equiv\int \rho\HH\rho\;d\rr=
   \left( \frac{3\lambda}{w}+V^{[0]} \right)
    \frac{e^{2D}}{(2\pi w)^{3/2}}
\eeq
where
\beq
V^{[n]}\equiv ({2\over \pi w})^{3/2}\int (\rr-\mm)^{n} V(r)
       e^{-2(\rr-\mm)^{2}/w} d\rr
\eeq
and 
\beq
N\equiv \int \rho \rho' d\rr =
[\pi (w+w')]^{-3/2}
  \exp\left\{-(\mm-\mm ')^{2}/(w+w')\right\}\exp( D+D') \quad.
\eeq
From Eq. \ref{matrix}, the  equations for  the variational
parameters are,
\BEN
\dot{w} &=& 4 \lambda  + 2 w V^{[0]} - \frac{8}{3} V^{[2]}\\
\dot{\mm} &=& - 2 {\bf V}^{[1]}\\
\dot{D} &=& \frac{1}{2} V^{[0]} - \frac{2}{w} V^{[2]}\quad\;.
\EEN
In absence of a potential, the exact free particle density matrix is recovered. 
The harmonic oscillator case is also correct since the Gaussian
approximation is exact there. 
For a hydrogen atom, $\lambda=1/2$,  $V(r)=-1/r$ and
\BEN
V^{[0]}&=&-\frac{1}{m}\erf\left(m\sqrt{2/w}\right)\\
{\bf V}^{[1]}&=&
{\mm\over m^3}{w\over 4}\left[ \erf\left(m\sqrt{2/w}\right)-
\sqrt{8\over \pi w}e^{-2m^2/w}\right ]\\
 V^{[2]}&=&\sqrt{w\over 2\pi}e^{-2m^2/w}+
    {3 w\over 4}V^{[0]}\quad.
\EEN
At low temperature, the density matrix as a function of $\rr$ goes to the ground state
wave function as discussed in more detail in the next section.
 One expects this to be a fixed 
point of the dynamics of the parameters $\mm$ and $w$ determined by 
$\dot{\mm}=0$ and $\dot{w}=0$ while $\dot{D}=-E_0$. 
The $\beta\rightarrow \infty$ fixed point: $\mm=0$, $w=9\pi/8$,
$\dot{D}=4/3\pi$ (atomic units) 
corresponds to the well known Rayleigh-Ritz variational result 
for a Gaussian trial wave function
\beq
    \label{eqrrh}
\Psi_0(\rr)=\left(4\over 3\pi\right)^{3/2}\exp(-8 r^{2}/9\pi)\;.
\eeq
In ground state variational studies, addition of two more Gaussians
brings the ground state energy to within $0.6$\% of exact and similar
improvement would be obtained here.

Results at finite $\beta$ require a numerical solution, which is
illustrated in the figure below comparing the Gaussian variational density matrix
with the exact \cite{Po88} and the free particle density matrix at several temperatures
for the initial condition $\rr'=1$. At high temperatures
($\beta=.05$ and $\beta=.25$)
the Gaussian approximation correctly reproduces the limiting free particle
density matrix. At lower temperatures, the cusp in the exact density matrix
due to the Coulombic singularity at the proton becomes evident and
the peak shifts to the origin somewhat faster than the Gaussian variational
approximation. As $\beta$ increases the exact result grows faster than
the variational since the correct energy, -0.5, is lower than $-4/3\pi$ but 
the Gaussian variational approximation remains rather accurate for $r>1$.
The free particle density matrix remains centered at $\rr=1$ and
beyond $\beta=0.5$ ($T=54.4$~eV) bears little resemblance to the
correct result.

\begin{figure}
\centerline{\psfig{figure=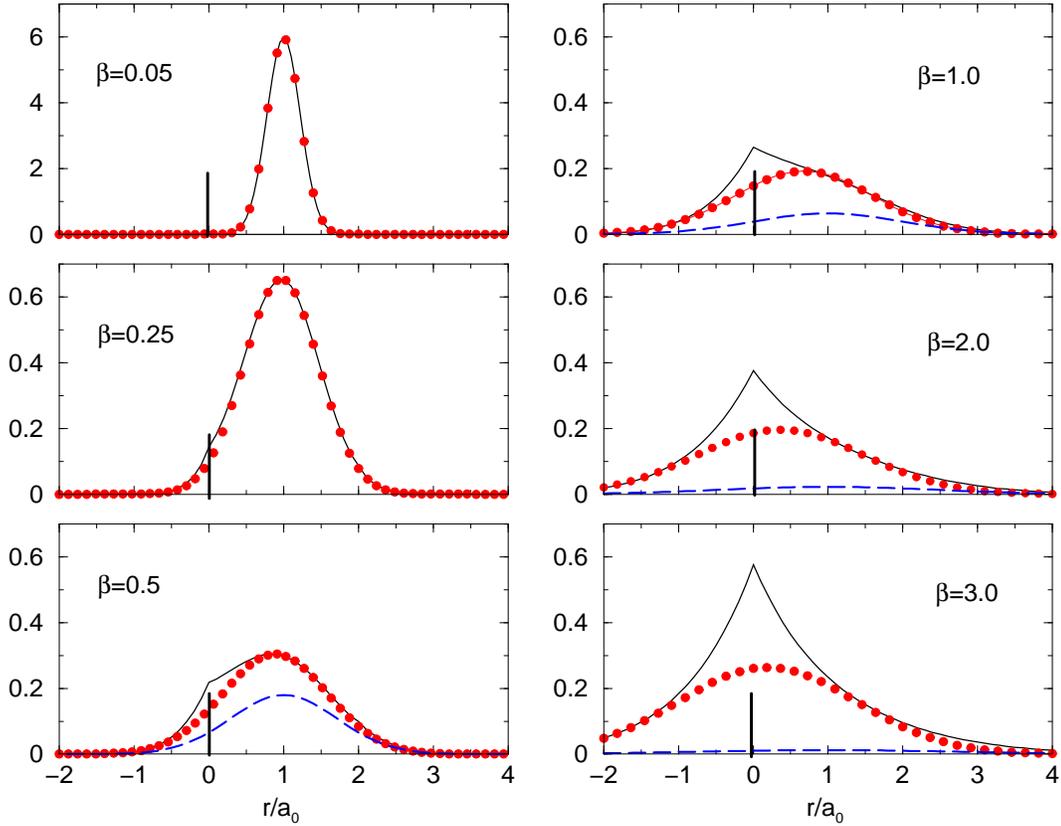,width=14cm,angle=270}}
\caption{ Comparison of the Gaussian variational approximation (circles)
 with the exact density matrix $\rho(\rr,\rr';\beta)$ (solid line) 
 for a hydrogen atom. The free particle density matrix (dashed line)
is also shown.  The plotted $r$ is along the line from the proton at 
the origin (marked by the vertical bar) through the initial electron 
position $\rr'=1$. }
 \label{FIG1}
\end{figure}

\section{Variational Density Matrix Properties}
\subsection{Zero Temperature Limit}
\label{ZeroTLimit}
      In the preceding section, it was shown that for the hydrogen atom the
Gaussian variational density matrix, as a function of $\RR$ converges at low
temperature to the Gaussian ground state wave function given by the Rayleigh-Ritz
variational principle. It is generally true that the Rayleigh-Ritz ground state
corresponds to a $\beta\rightarrow\infty$ of the variational density matrix as we
now show.

      The  Rayleigh-Ritz principle states that for any real parameterized wave function
$\Psi(\RR,q_{1},\ldots,q_{m})$ the variational energy
\beq
      \label{eq39}
E(\{q\})=\begin{array}{c}\underline{\int  \psi(\RR)\HH \psi(\RR) \;d\RR}\\
\int  \psi(\RR)^{2} \;d\RR\end{array}
\eeq
is greater than or equal to the true ground state energy even at the minimum determined
by
\beq
      \label{eq40}
\begin{array}{c}\partial\\\overline{\partial q_{k}}\end{array} E(\{q\})=0\;\;\forall k\;.
\eeq
For the VDM ansatz, an amplitude parameter $D$ is assumed such that
\beq
\rho(\RR,\RRp;\beta)=e^{D(\RRp;\beta)}\tilde{\rho}(\RR,\{q(\RRp;\beta)\})\;.
\eeq
As in the one particle example, it is expected that at low temperature, $\beta\rightarrow\infty$,
the other $\dot{q}_{k}\rightarrow 0$ while $\dot{D}\rightarrow$ constant.
From this assumption, Eq. \ref{matrix} implies that as $\beta\rightarrow \infty$ 
\BEN
      \label{eq42}
\frac{\partial H}{\partial q_k} + \dot{D} \frac{\partial N}{\partial q_k} = 0
\EEN
for all variational parameters, where we have defined
$H\equiv\int \rho \HH \rho\;d\RR$ and $N\equiv\int \rho^{2}\;d\RR$.
      Since $\partial H/\partial D=2H$ and  $\partial N/\partial D=2N$, Eq.~\ref{eq42} 
for $q_{k}=D$ implies $\dot{D}=-H/N\equiv -E_{0}$ so Eq.~\ref{eq42} may be rewritten as
\BEN
\frac{\partial }{\partial q_k} \left( \frac{H}{N} \right) = 0
\EEN
at the $\beta\rightarrow \infty$ fixed point. With the correspondence
\beq
      \label{eq44}
\rho(\RR,\{q(\RRp,\beta)\})\rightarrow e^{D(\RRp;\beta)}\psi(\RR,\{q\})\quad,
\eeq
this is equivalent to Eq. \ref{eq40} and thus the Rayleigh-Ritz ground state
corresponds to a zero temperature fixed point in the dynamics of the
parameters.

$D$ is a function of $\RRp$ and $\beta$, which is calculated by 
integrating from $\beta=0$ with Eq. \ref{fpl} as initial conditions. 
The zero temperature limit of $\dot{D}$ is a constant, $-E_0$, which means in the
low temperature limit
$D$ can written as
\beq
D(\RRp;\beta) = - \beta E_0 + f(\RRp)\quad.
\eeq
The function $f(\RRp)$ can be rewritten as,
\beq
f(\RRp) = \ln \left\{ \psi_0(\RRp) \left[ \, 1 + \delta(\RRp) \, \right] \right\}\quad,
\eeq
where the function $\delta(\RRp)$ is introduced to describe the variational error in
the solution of the Bloch equation. It is identical to zero if the variational 
ansatz includes the exact solution. It leads to loss of symmetry in 
$\RR$ and $\RRp$, which will discussed in the next section. 
Eq.~\ref{eq44} now reads,
\beq
    \label{eq47}
\rho(\RR,\RRp,\beta\to\infty) = e^{-\beta E_0} \psi_0(\RR) \psi_0(\RRp)
\left[ 1 + \delta(\RRp) \right]
\label{zeroRho2}
\eeq

For certain potentials, several fixed points of the dynamics can exist. 
From Eq. \ref{eq47}, it follows that only the lowest energy state contributes to 
physical observables calculated from Eq. \ref{thermal_average}. This completes the argument 
that the zero temperature limit of the VDM correspond to the Rayleigh-Ritz ground state.

In case of an anti-symmetrized ansatz for the density matrix, one can show
that the fixed point of the dynamics in imaginary time corresponds to the 
Rayleigh-Ritz ground state for an anti-symmetrized wave function.

\subsection{Loss of Symmetry}
\label {loss}

      The exact density matrix is symmetric under $\RR\leftrightarrow\RRp$.
Since we have singled out $\RRp$ as the initial point for the imaginary
time dynamics, it is not clear that the approximation given in 
Eq.~\ref{gauss} automatically satisfies this condition. For the free particle limit
and the harmonic oscillator, where the Gaussian is the exact solution, it obviously
does but in general it does not.

    As a specific example, consider again the ground state limit of the  
hydrogen atom where the Gaussian VDM approximation. Eq. \ref{gauss} 
then reads,
\beq
\lim_{\beta\rightarrow\infty}\;\rho(\rr,\rr ';\beta) = 
e^{D(r';\beta)} \; (8/9\pi^{2})^{3/2} e^{-8r^{2}/9\pi}\;.
\eeq
For this to be symmetric under $\rr\leftrightarrow \rr '$, we must have
\beq
         \label{eq49}
 \lim_{\beta\rightarrow\infty} D(r';\beta)=-8 r'^{\,2}/9\pi + c(\beta)
\eeq
and from the result for $\dot{D}$, $\lim_{\beta\rightarrow\infty} 
c(\beta) =  4\beta/3\pi +c_{1} $.

\begin{figure}
\centerline{\psfig{figure=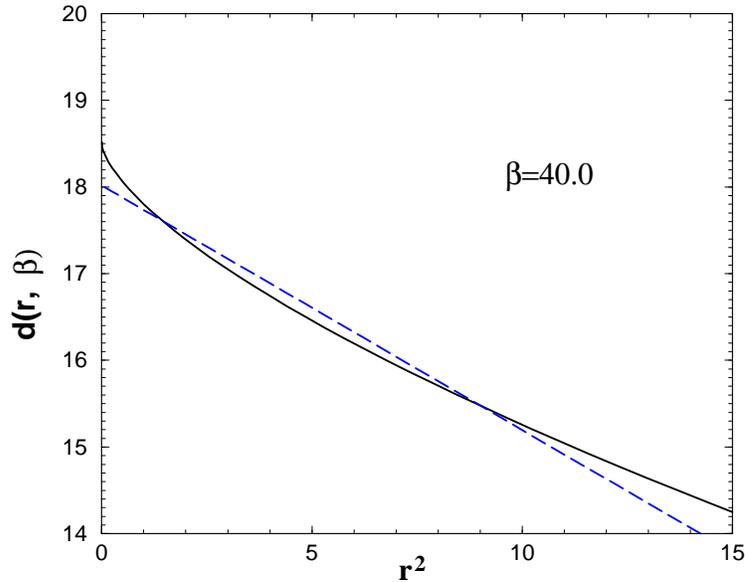,width=10cm,angle=0}}
\hspace*{1mm}
 \caption{ $D(r,\beta)$ from the Gaussian approximation in the ground state limit 
          (solid line) of the hydrogen atom. Deviations of this function from linearity 
          indicate a breakdown of symmetry in the Gaussian approximation for
          $\rho(\rr,\rr ';\beta)$. The dashed line is 
$-8r^{2}/9\pi+4\beta/3\pi+3/2\ln 2$ expected from the Rayleigh-Ritz
ground state Eq.~\ref{eqrrh}.}
 \label{FIG2}
\end{figure}

       Figure \ref{FIG2} compares the $D(r,\beta)$ from the Gaussian 
VDM with Eq. \ref{eq49} using \mbox{$c(\beta)=4\beta/3\pi+3/2\ln 2$.}

           There are several consequences of this small violation of 
$\RR\leftrightarrow\RRp$ symmetry. As shown generally in the section 
above, in the $\beta\rightarrow\infty$ limit $ -\dot{D}$ is the 
Rayleigh-Ritz variational
ground state energy for a Gaussian wave function, which for the hydrogen atom is 
$E_0=-4/3\pi = -0.4244$. Because of
the loss of symmetry this is not the same as the energy given by the estimator
\beq
   \label{Eestimator}
   \left<E\right> = \left<\HH\right> \equiv \frac{\Tr[\HH\rho]}{\Tr[\rho]}
\eeq
in the $\beta\rightarrow \infty$ limit,
which for the hydrogen atom gives the more accurate result $\left<E\right>=-0.4709$.
This will be seen again below for the hydrogen molecule where 
Eq. \ref{Eestimator} also gives more accurate ground state energies.
Other consequences are less pleasant. Although the energy is more accurate
the virial theorem, $\left<K\right>=-\left<U\right>/\,2$, 
between the kinetic and potential energy
is violated by about $3\%$ (while both are more accurate than the
usual ground state variational Gaussian result). This has consequences for 
calculating the equation of state particularly at low density.
Slightly more complicated, explicitly symmetric forms for the VDM could be used but in this
paper we will continue to explore the basic Gaussian approximation.

\subsection{Thermodynamic Estimators}

     Since the VDM, except in the simplest cases,
is not exact
various estimators for the same quantity will differ. For example the variational 
principle introduced in section II consists essentially in globally minimizing the
squared difference between $\partial\rho/\partial\beta$ and $\HH\rho$, either of
which can be used in estimating the energy. As mentioned above the energy estimator
Eq. \ref{Eestimator} and its kinetic and potential energy pieces do not automatically
satisfy the virial theorem for Coulomb systems at low density.
   As an alternative to Eq. \ref{Eestimator}, one can use the thermodynamic estimators,
\BEN
\left< E \right> &=& - \left< \frac{\partial}{\partial \beta} \ln \rho \right>,
\label{thermoE}\\
\left< T \right> &=& - \frac{\lambda}{\beta} \left< \frac{\partial }{\partial \lambda}
             \ln \rho \right>,
\label{thermoT}\\
\left< V \right> &=& - \frac{e^2}{\beta} \left< \frac{\partial}{\partial e^2}  \ln \rho \right>
\label{thermoV}
\EEN
for the total, kinetic and potential energy. These estimators satisfy 
\beq
\left< E \right>=\left< T \right>+\left< V \right>
\label{thermo}
\eeq
by the following argument. Any function $f=f(\beta \lambda,\beta e^2)$ satisfies
\beq
\beta \frac{\partial f}{\partial \beta} = 
\lambda \frac{\partial f}{\partial \lambda} +
e^2 \frac{\partial f}{\partial e^2}\quad.
\eeq
From Eq. \ref{matrix} it follows that all parameters $q_i=q_i(\RRp;\beta,\lambda,e^2)$ have 
this property and therefore so does the variational density matrix.

In the zero temperature limit, the thermodynamic estimators satisfy the virial theorem,
which is also satisfied by any exact and any variational Rayleigh-Ritz ground state.
From the zero temperature limit of the VDM given by Eq. \ref{zeroRho2} and the $1/\beta$
factor in Eqs. \ref{thermoT} and \ref{thermoV}, it is seen that the symmetry error
$\delta(\RRp)$ is unimportant in this limit.
It should be noted that 
calculating the derivatives for $\left< T \right>$ and $\left< V \right>$ increases the 
numerical work.
The pressure is estimated from
\beq
3\, \left< P \right> v= 2 \left< K \right> + \left< V \right>\;.
\eeq

\section{Many particle density matrix}
  We represent the many particle density matrix by a determinant of 
one-particle density matrices (Eq. \ref{eq:1.3}). It can written as,
\beq
\label{product}
\rho(\RR,\RRp,\beta) =\sum_{\cal{P}}\epsilon_{\cal{P}} 
                   \prod_{k} \rho_1(\rr_k,\rr'_{{\cal{P}}_k},\beta) = 
\sum_{\cal{P}}\epsilon_{\cal{P}} e^D \prod_{k} \:(\pi w_{{\cal{P}}_k})^{-3/2} \:
       \mbox{exp} \left\{ -\frac{1}{w_{{\cal{P}}_k}} (\rr_k-\mm_{{\cal{P}}_k})^2
            \right\}\;.
\eeq
 The permutation sum is over all permutations
of identical particles (e.g. same spin electrons) and the permutation 
signature $\epsilon_{\cal{P}}=\pm 1$.
The initial conditions for Eq. \ref{matrix}
are $w_k=0$, $\mm_k = \rr_k'$, and $D=0$. 
For this ansatz the generator of the norm matrix, Eq. \ref{norm_gen},
\beq
                       \label{eq4.2}
N= \exp( D+D') \, \sum_{\cal{P}}\epsilon_{\cal{P}}\prod_{k} 
[\pi (w_{k}+w_{{\cal P}_k}')]^{-3/2}
  \exp\left\{-(\mm_{k}-\mm_{{\cal P}_k} ')^{2}/(w_{k}+w_{{\cal P}_k}')\right\}
\;.
\eeq
For a periodic system the above equation is also summed over all periodic
simulation cell vectors, ${\bf L}$, with $\mm_{k}-\mm_{{\cal P}_k}\rightarrow
\mm_{k}-\mm_{{\cal P}_k}+{\bf L}$.
If only the identity permutation is considered the norm matrix
is easily inverted so that Eq. \ref{matrix} gives
\BEN
\label{wdot}
\dot{w}_k &=& - 2 w_k H_D -\frac{8}{3} w_k^2  H_{w_k} \\
\label{mdot}
\dot{\mm}_k &=& - w_k H_{\mm_k}\\
\label{ddot}
\dot{D}   &=& 
- \left( \frac{3}{2} n + 1 \right) H_D
- 2 \sum_{i=1}^n w_i H_{w_i}
\quad,\\
\label{Hq}
\mbox{where} \quad\quad H_{q_k} &=& \frac{1}{2}\frac{\partial H}{\partial q_k}\;.
\EEN
For systems of electrons and ions the full expression for $H_{q_k}$ and the norm
matrix are derived in Appendix A.

\begin{figure}[htb]
\centerline{\psfig{figure=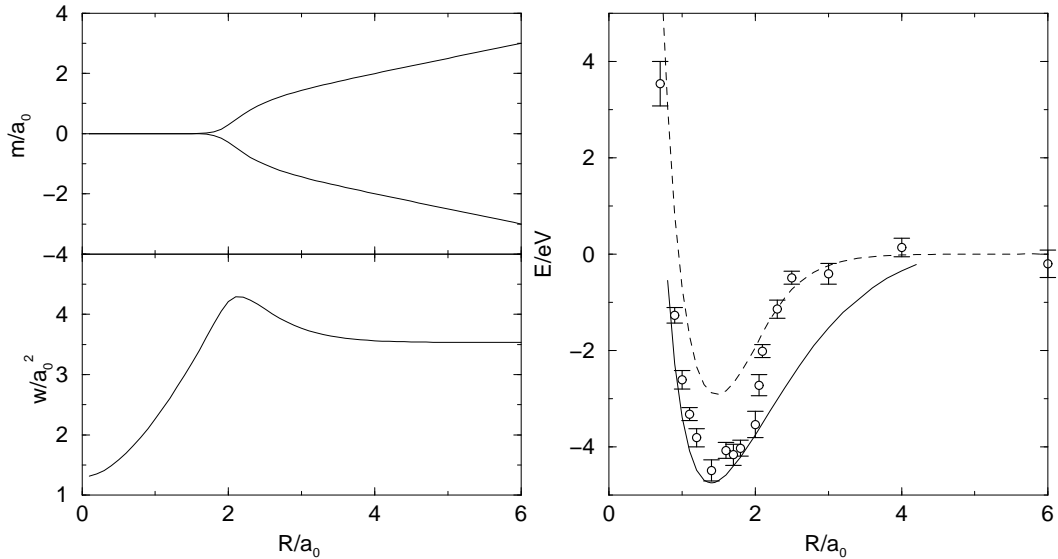,width=14cm,angle=-90}}
\vspace*{2mm}
\caption{ Gaussian approximation for the ground state of a hydrogen
           molecule for bond length $R$. The top left  panel shows the Gaussian
           mean parameter $\mm$ for the two electrons. These stay in 
           the center of the bond ($\mm=0$)  until about $R=2 a_0$ and then 
           attach themselves to the separating protons ($\pm\; R/2$). The width 
           parameter, displayed in the lower left panel, makes the transition from
           the optimal value for a helium atom, $R=0$, to the hydrogen atom
           result $w=9\pi/8 a_0^2$ at large $R$. The right panel shows the dissociation
           energy for the singlet state computed from 
           Eq. \ref{Eestimator} (open circles with error bars) and the 
          thermodynamic estimator ($-dD/d\beta$) (dashed line) compared
           to the results of Kolos and Roothan (solid line).}
\label{FIG3}
\end{figure}

Application to an isolated hydrogen molecule at low temperature is shown in Figure \ref{FIG3}.
This is for the singlet state (anti-parallel electron spins). The triplet state
is considered later after a discussion of how to treat permutation terms in the
parameter equations.
The bond length at minimum energy is 1.47~a$_0$, compared with the experimental value of 
 1.40~a$_0$. The direct energy estimator Eq.~\ref{Eestimator} gives a
dissociation energy of 4.50~eV at the minimum compared to the experimental value of
4.75~eV. Beyond $R=2$, the energy rises quickly toward the value
given by the Rayleigh-Ritz estimator $-dD/d\beta$. 

\section{Antisymmetry in the Parameter Equations}
      The determinantal form for the VDM, Eq. \ref{product}, is
correctly antisymmetric under exchange of identical particles. Since ion exchange effects
are negligible at the temperatures considered here these are ignored.

      The determinantal form leads to $N!$ terms in the equations of motion for the
variational parameters presented in appendix A. It was originally hoped that exchange
effects could be ignored in these equations while retaining the full determinantal form
for the VDM but this leads to an instability in fermionic systems, e.g. it results
in an unphysical strong attraction between two hydrogen molecules.

A practical means of treating all exchange terms, in particular terms involving the
potential energy, in the variational parameter equations was not found.
Instead it was necessary to use an approximation similar to that used in the
real time computations \cite{KTR94b,KRT98}: only
pair exchanges in the kinetic energy terms were retained. This will be illustrated
for the hydrogen molecule after first giving the explicit form for this correction.
It is stressed that, unlike the real time computations, once the
variational parameters are determined the full determinantal
form is then used in calculating the various averages.

\begin{figure}
 \leavevmode
\centerline{\psfig{figure=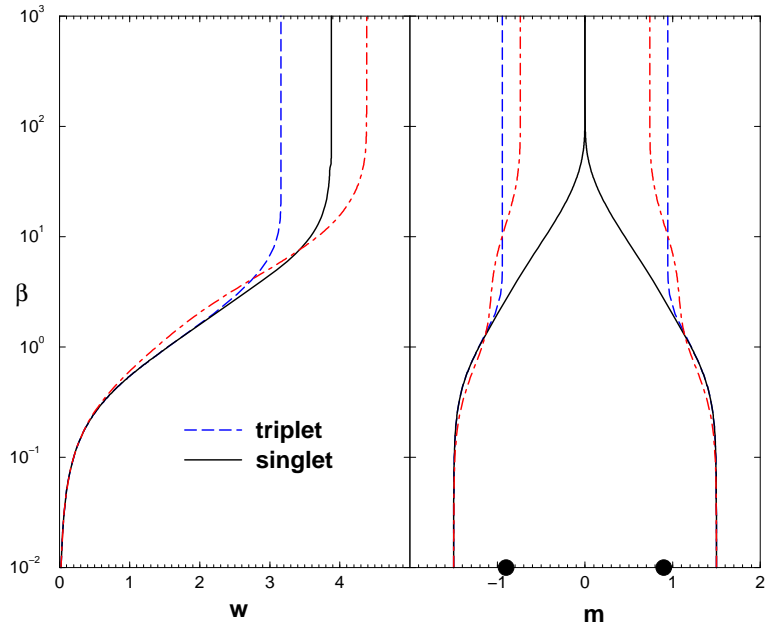,width=10cm,angle=270}}
\vspace*{2pt}
\caption{ Effect of antisymmetry on the density matrix parameters, width and mean, for a 
          hydrogen molecule. The protons (large black dots along x axis) are separated by
          $1.8 a_0$ and the initial electron positions $r_e(\beta=0)=\pm 1.5 a_0$ along
          the molecular axis. The solid line for the singlet state (electron spins anti-parallel)
          shows both electrons centered in the molecular bond at low temperatures 
         (large $\beta$). For the triplet state (parallel electron spins), long
          dashed line the electrons are centered close to the protons. The approximation
          of including only kinetic pair exchanges (dot-dashed line) gives a similar
          result for the mean, with the electrons centered slightly inside the protons but
          overestimates the Gaussian width (left panel). At high temperature ($\beta\leq 4$)
          exchange is unimportant and the parameters are nearly the same for all cases.}
\label{H2fig}
\end{figure}

For two particles with parallel spin, the correction term to the kinetic energy 
is given by,
\BEN
\Delta T &=& \frac{N_I}{N_{AS}} \int \! d\RR \;
\rho_{AS} \; \hat{T} \; \rho_{AS} \quad-\quad 
\int \! d\RR \;
\rho_{I} \; \hat{T} \; \rho_{I}\\
\rho_{AS} &=& \rho_1(\rr_1)\rho_2(\rr_2) - \rho_2(\rr_1)\rho_1(\rr_2) \quad, \quad
\rho_{I} = \rho_1(\rr_1)\rho_2(\rr_2)\\
N_{AS} &=& \int \! d\RR \; \rho_{AS}^2 \quad,\quad 
N_{I} = \int \! d\RR \; \rho_{I}^2
\EEN
For the Gaussian ansatz in Eq. \ref{product} it becomes,
\BEN
\Delta T &=& - 
\frac{4 \lambda N_I }{w N_Q} 
\left[ \, 3 \left(1-\wt^2\right)-Q^2 \right]
\quad,\\
w &=& w_1+w_2 \quad,\quad 
\wt = \frac{w}{2\sqrt{w_1 w_2}}\quad,\quad
Q^2 = \frac{2}{w} \left( \mm_1-\mm_2 \right)^2 \quad,\quad 
N_Q = \wt^3 e^{Q^2} - 1\quad.
\EEN
The corrections to the norm matrix $\NN$ are neglected in order to keep its analytically
invertible form. 
      The corrections to $H_{q_k}$ in Eq. \ref{Hq} are given by 
\BEN
\Delta T_{q_k} = \frac{1}{2 N_I} \frac{\partial}{\partial {q_k}} \Delta T
\EEN
The correction to dynamics of the parameters follow from Eq. \ref{wdot} to \ref{ddot},
\BEN
\Delta \dot{w}_1 &=& -2 \, w_1 
\left( \Delta T_{D} + \frac{4}{3} w_1 \: \Delta T_{w_1} \right)\\
\Delta \dot{\mm}_1 &=& - w_1 \: \Delta T_{\mm_1}\\
\Delta \dot{D} &=& -2 
\left( \Delta T_{D} + w_1 \: \Delta T_{w_1} + w_2 \: \Delta T_{w_2} \right)\quad.
\EEN
These equations lead to an effective repulsion between the Gaussians 
for two electrons with parallel spin if there is significant overlap.
       As a example of this effect the variational parameters for the
 singlet and triplet states of the
hydrogen molecule are compared in Fig. \ref{H2fig}.
For the triplet state parameters the solution including full exchange effects (long
dashed line) are compared with those obtained in the kinetic pair exchange approximation
 (dot-dashed line). The approximation now prevents the Gaussian means for the same spin
electrons from collapsing to the bond center at lower temperature and is
numerically close to the solution for full exchange.
      Even at the lowest temperature considered here in the dense hydrogen simulations ($5000$ K)
exchange effects between same spin electrons are negligible beyond a 
few angstroms, i.e.  one or perhaps two nearest neighbors. 
Fig. \ref{H2fig} for the triplet state thus
overestimates the effect likely in dense hydrogen. The main effect of including exchange
in the parameter equations is probably to prevent the instability mentioned above.

\begin{figure}
 \leavevmode
\centerline{\psfig{figure=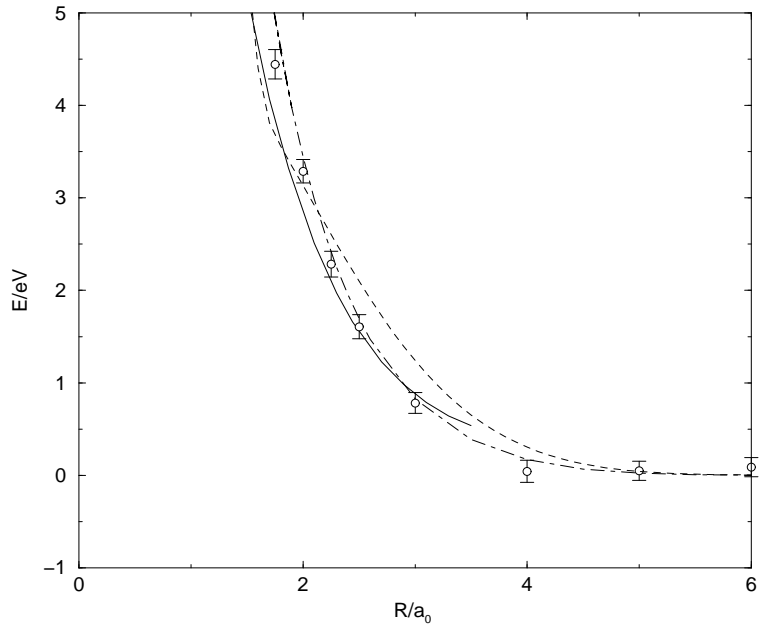,width=10cm,angle=270}}
\caption{ Energy of repulsion for the triplet ground state of the
          hydrogen molecule for bond length $R$. 
	The thermodynamic (dashed line) and the direct estimator, Eq. 
       \ref{Eestimator}, (circles with error bars) for the 
        Gaussian approximation using the kinetic exchange term
        in the parameter equations  
        are compared with the Kolos and Roothan results (solid line).
        The thermodynamic estimator
        for the Gaussian approximation with all exchange terms is shown by the dot-dashed line.}
\label{H2tri}
\end{figure}

      Fig. \ref{H2tri} shows an energy comparison for the triplet 
ground state of the hydrogen molecule.
First, we compare the Gaussian approximation using only the kinetic 
exchange term in the parameter equations. 
For the direct estimator, Eq. \ref{Eestimator}, one finds fairly good 
agreement with the quantum chemistry result \cite{KR69}.
The thermodynamic estimator gives a somewhat more repulsive
triplet interaction for $R>2a_0$.  Considering also 
the Coulomb exchange terms in the Gaussian approximation leads to 
the dot-dashed line for the thermodynamic
estimator. We conclude that leaving out the Coulomb exchange terms in 
the parameter equations for efficiency reasons is a reasonable 
approximation in many particle simulations.

\section{Results from many particle simulations}

    In this section, we report results from VDM Monte Carlo simulation with 32 pairs of protons 
and electrons in the temperature and density range of $5\,000\,$K$\,\leq T \leq 250\,000\,$K 
and $1.75 \leq r_s \leq 4.0$. Although the Gaussian ansatz VDM will be 
seen to provide a reasonable model for hydrogen over the full density 
and temperature regime, a large purpose in presenting these results is 
to serve as a base for documenting future improvements
from better VDMs and the application of RPIMC. 

\begin{figure}[htbp]
\centerline{\psfig{figure=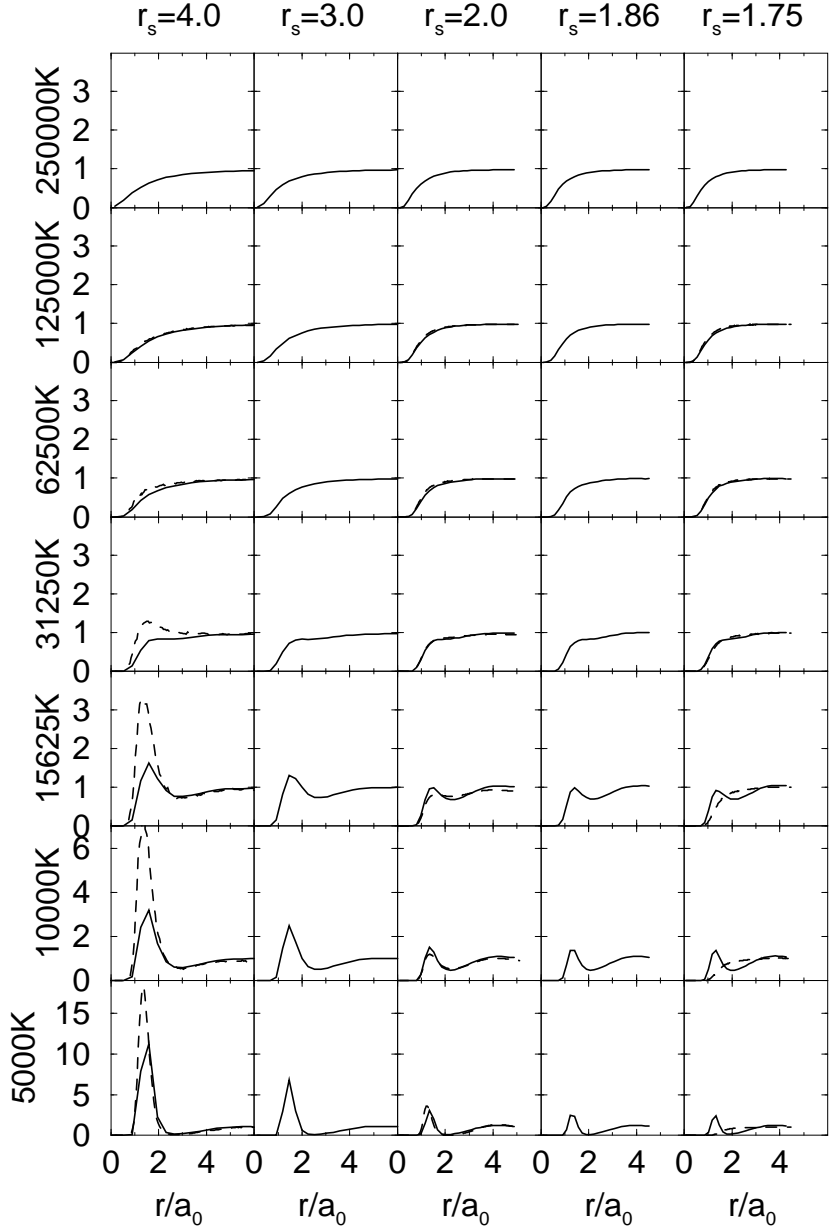,width=11cm,angle=270}}
\caption{ Proton-proton pair correlation function from VDM (solid line) 
	  and RPIMC (dashed lines at $r_s$=1.75, 2.0, and 4.0 for $T \leq 125\,000\,$K).}
\label{grpp}
\end{figure}

The proton-proton pair correlation functions are shown in Fig. \ref{grpp}. 
For temperatures below $20\,000\,$K, a peak emerges near $1.4 a_0$ that demonstrates 
clearly the formation of molecules. The comparison with RPIMC simulations \cite{Ma96,MC99} at low density
shows that the peak positions agree well but RPIMC predicts a 
significantly bigger height indicating a larger number of molecules. This could be explained by the missing correlations 
in the VDM ansatz.

At a density of $r_s=2.0$, proton-proton pair correlation functions from RPIMC and VDM 
are almost identical. The area under the peak multiplied by the density gives an estimate 
for the molecular fraction. By comparing the estimate for different densities one finds that 
the molecular fraction is diminished when the density is lowered below $r_s=2.0$.
This effect is well-known and is a result of the increased entropy of dissociated molecules.

Considerable differences between the proton-proton pair correlation 
functions are found at 
$r_s=1.75$ below $T=20\,000\,K$ where VDM shows still a fair number of molecules while 
RPIMC predicts a metallic fluid where all bonds are broken as a 
result of pressure dissociation \cite{Ma96,Mi99}. This effect has to be verified by RPIMC 
simulations with VDM nodes because free particle nodes could enhance the transition to a 
metallic state.

The peak positions shifts from 
$1.45 a_0$ at a low density of $r_s=4.0$ to $1.3 a_0$ at $r_s=1.75$. 
The same trend has been found in the RPIMC simulations \cite{Ma96} but the opposite was 
reported in \cite{Ga99,Re99}.

\begin{figure}[htbp]
\centerline{\psfig{figure=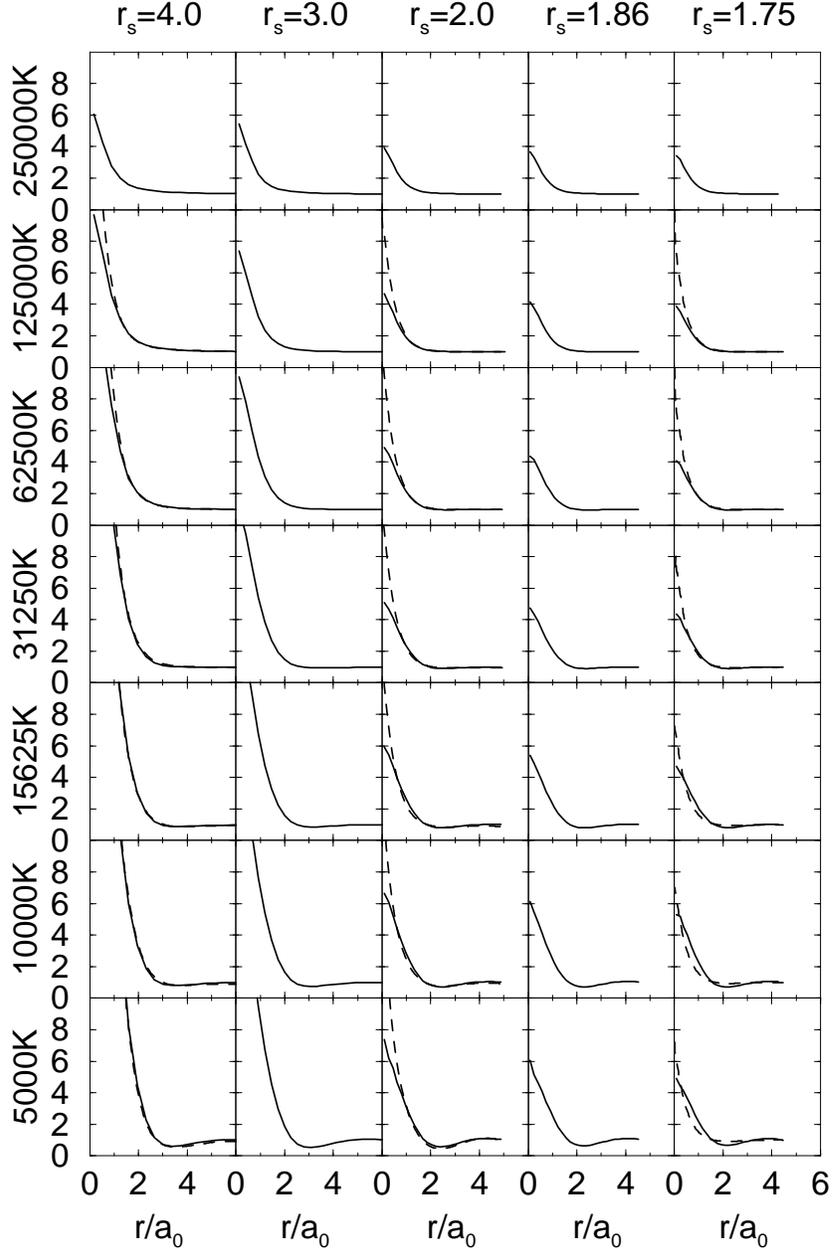,width=11cm,angle=270}}
\caption{ Proton-electron pair correlation functions from
          VDM (solid line) and RPIMC (dashed lines at
          $r_s$=1.75, 2.0, and 4.0 for $T \leq 125\,000\,$K).   }
\label{grpe}
\end{figure}

In the proton-electron pair correlation functions shown in Fig. \ref{grpe}, one finds a 
strong attraction present even at high temperatures such as $250\,000\,$K.
At low temperatures, the electrons are bound in atoms and molecules.
This pair correlation function does not show a clear distinction between the two cases.
From studying the height of the peak at the origin multiplied by the density, one can
estimate the number of bound states at low temperature. Similar to the molecular fraction
one finds a reduction of bound electrons with decreasing density below $r_s=2.0$. The comparison
with PIMC shows that VDM underestimates the height of the peak.  This 
is probably a result of the Gaussian ansatz, which does not satisfy 
the cusp condition at the proton. 

\begin{figure}[htbp]
\centerline{\psfig{figure=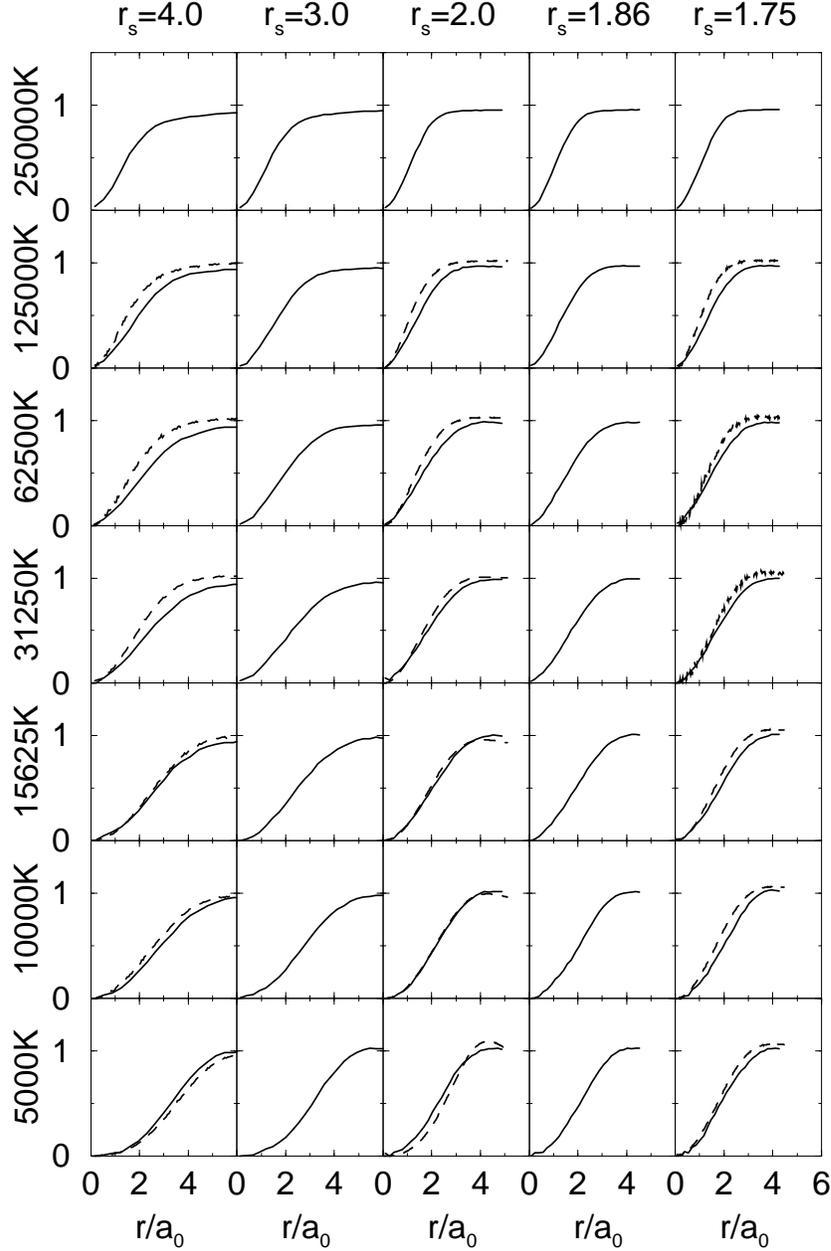,width=11cm,angle=270}}
\caption{ Electron-electron pair correlation function for electron with parallel spin from
          VDM (solid line) and RPIMC (dashed lines at
          $r_s$=1.75, 2.0, and 4.0 for $T \leq 125\,000\,$K). }
\label{greel}
\end{figure}

\begin{figure}[htbp]
\centerline{\psfig{figure=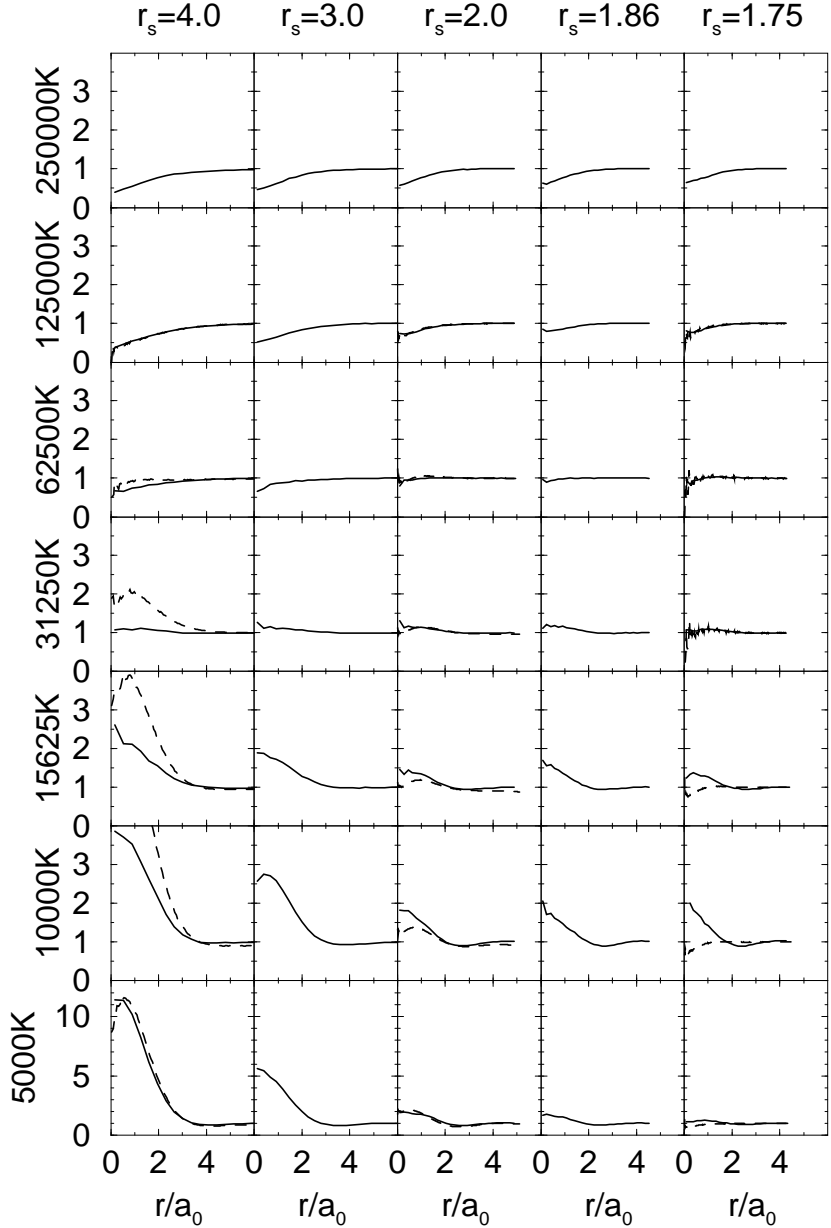,width=11cm,angle=270}}
\caption{ Electron-electron pair correlation function for electron with anti-parallel spin from
          VDM (solid line) and RPIMC (dashed lines at
          $r_s$=1.75, 2.0, and 4.0 for $T \leq 125\,000\,$K).   
          Note the change in scale in the last row.}
\label{greeu}
\end{figure}

Fig. \ref{greel} shows the effect of the Pauli exclusion principle leading the a strong 
repulsion for electrons in the same spin state. This effect is not present in the interaction
of electrons with anti-parallel spin (Fig. \ref{greeu}). At high temperature, one observes
the effect of the Coulomb repulsion. At low temperature, one finds a peak at the origin that
is a result of the formation of molecule, in which two electrons of opposite spin are
localized along the bond. The differences to the PIMC graphs can be interpreted as a consequence of
different molecular fractions, which has also been observed in Fig. \ref{grpp}.

\begin{figure}[htbp]
\centerline{\psfig{figure=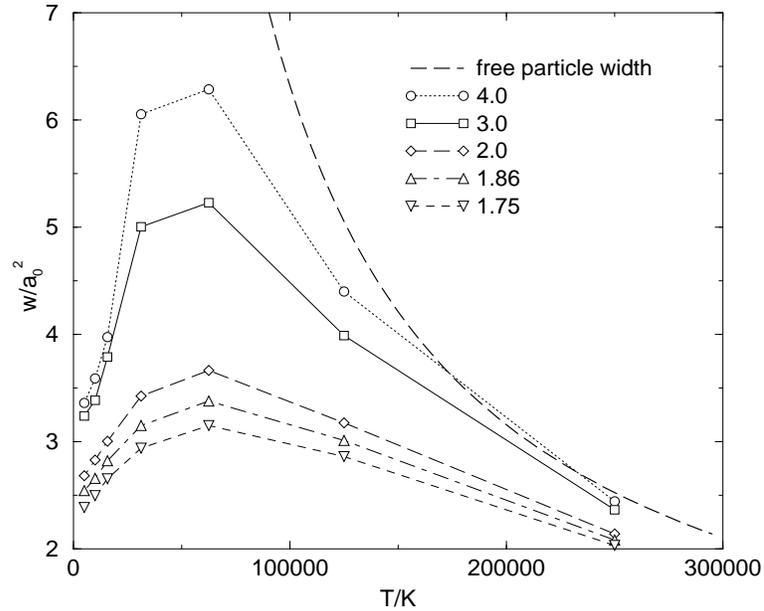,width=10cm,angle=0}}
\caption{ Average width of the Gaussian single particle density matrices as a function of temperature 
for different densities}
\label{averageWidth}
\end{figure}

The average width $w$ of the Gaussian is shown in Fig. \ref{averageWidth} as a function 
temperature and density. At high temperature and low density, one finds
only small deviations from the free particle limit. 
These become more significant with increasing density and
decreasing temperature. At low temperature, the attraction to the
protons dominates, which leads to a decreasing average width. Finally bound states form
and the width approaches a finite limit. At low densities, this is close to
the ground state width of the isolated molecule $3.138\,a_0^2$.

\begin{figure}[htb]
\centerline{\psfig{figure=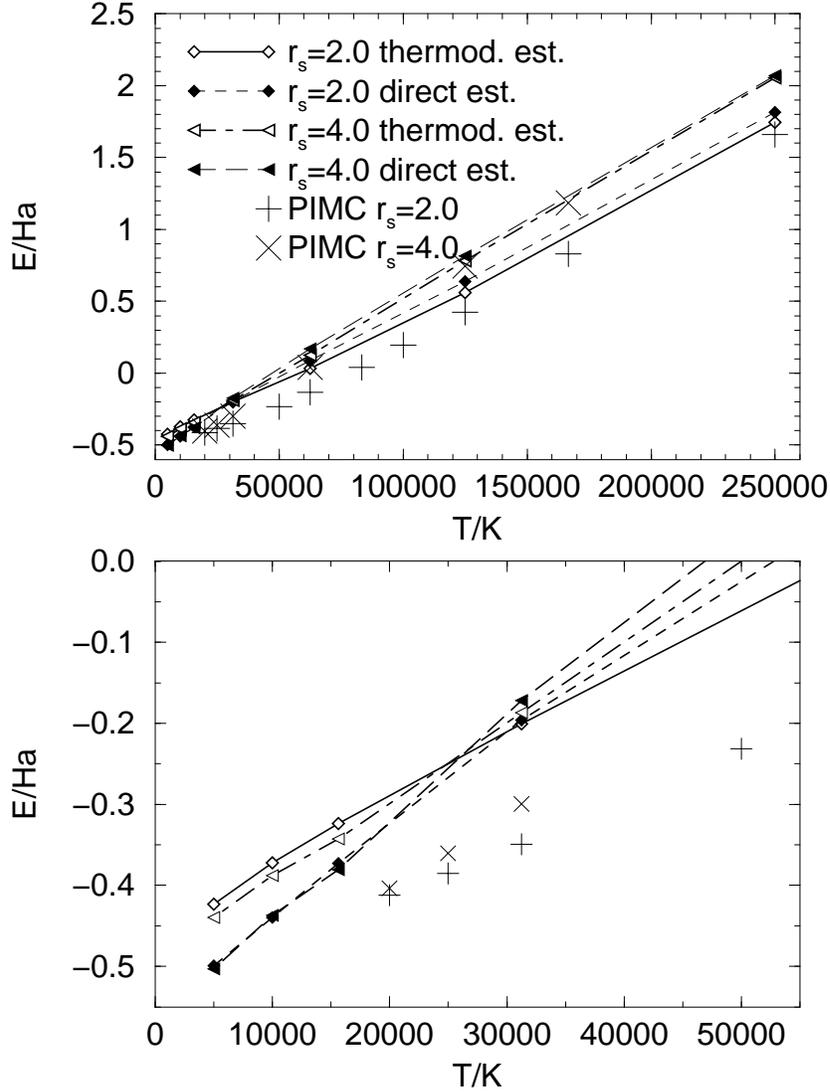,width=11cm,angle=270}}
\caption{ Internal energy per atom versus temperature}
\label{E_vs_T}
\end{figure}

   In Fig. \ref{E_vs_T}, we compare the internal energy from the thermodynamic estimator in
Eq. \ref{thermoE} and the direct estimator \ref{Eestimator}. Both agree fairly well at low 
density. Differences build up with increasing density and decreasing temperature.
Comparing with RPIMC simulations, one finds that the VDM energies are generally too high.
The magnitude of this discrepancy shows the same dependence on density and temperature like 
the difference between the two VDM estimators. The difference to the RPIMC results could be 
explained by the missing correlation effects in the VDM method.

At high temperature, the thermodynamic estimator always gives lower energies 
than the direct estimator. Below $T=25\,000\,$K, the ordering is reversed.
This is consistent with the results from the isolated atom 
and molecule. The consequence is that the direct estimator is actually closer to the 
value expected from RPIMC simulations. However, it should be noted that this estimator is
not thermodynamically consistent (see section \ref{loss}).

\begin{figure}[htb]
\centerline{\psfig{figure=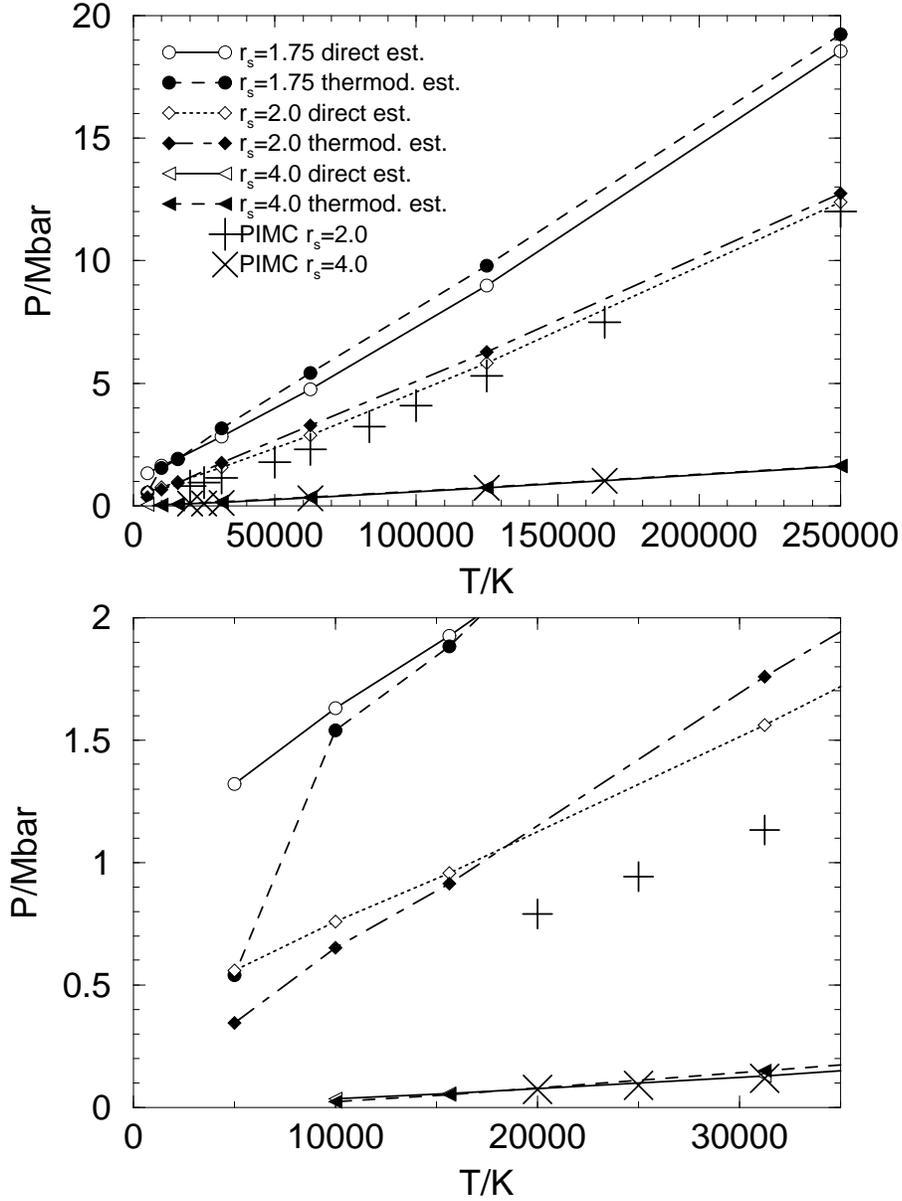,width=12cm,angle=270}}
\caption{ Pressure versus temperature in high and low temperature range.
VDM pressure is calculated from virial relation using both the direct and
thermodynamic estimators for kinetic and potential energy.}
\label{P_vs_T}
\end{figure}

In Fig. \ref{P_vs_T}, we compare pressure as a function of temperature
and density from the two VDM estimators with RPIMC results. At low density, the agreement
is remarkably good. With increasing density and decreasing temperature, the difference
grows. For densities over $r_s=2.0$ below $10\,000\,$K, one finds a significant drop in the
direct estimator for the pressure. We interpret this effect 
as a result of the thermodynamic inconsistency.

\begin{figure}[htb]
\centerline{\psfig{figure=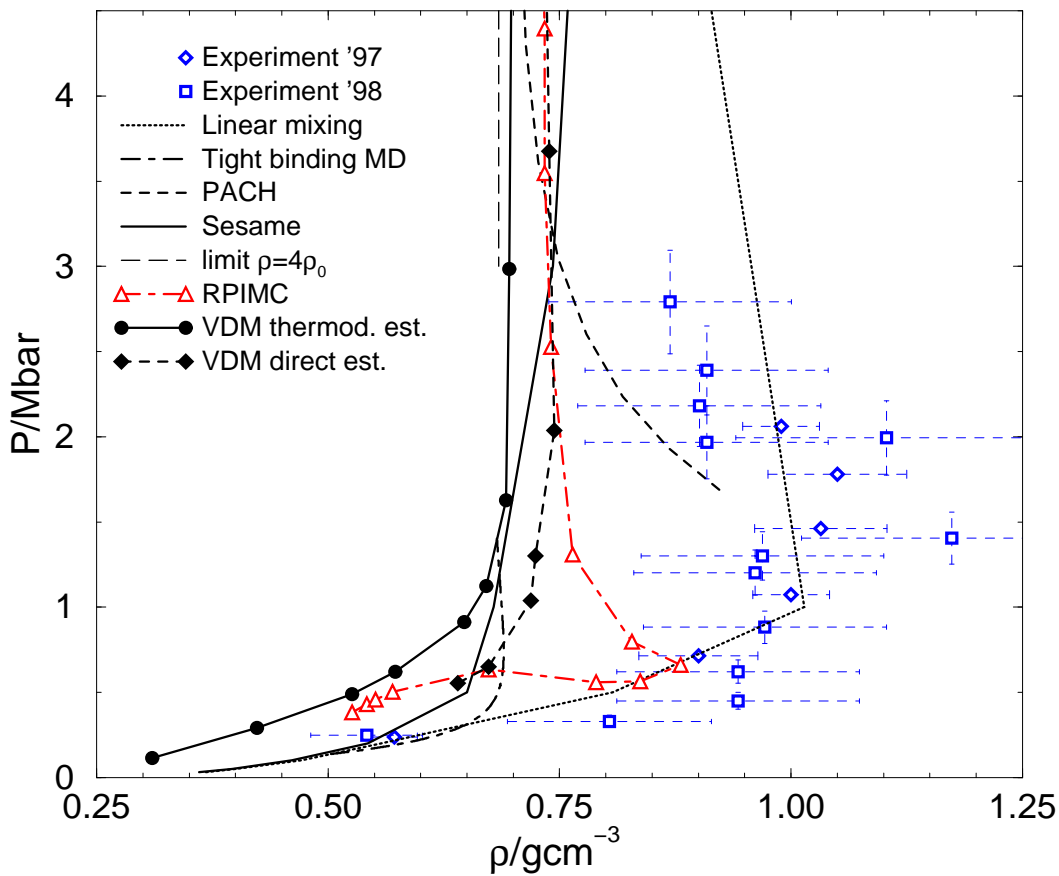,width=14cm,angle=270}}
\caption{ Comparison of experimental and several theoretical Hugoniot functions}
\label{hugoniotFigure}
\end{figure}

   Fig. \ref{hugoniotFigure},  compares the Hugoniot
from Laser shock wave experiments \cite{Si97,Co98} with results from 
several theoretical approaches 
(Sesame data base by Kerley \cite{Ke83} (thin solid line), 
linear mixing model by Ross (dashed line) \cite{Ro98}, 
tight-binding molecular dynamics by Lenosky {\it et.al.} \cite{Le97} (dash-dotted line), 
Pad\'e approximation in the chemical picture by Ebeling {\it et.al.} \cite{rotesbuch} (dotted line),
RPIMC simulations \cite{Mi98} (triangles), 
VDM direct estimator (full diamonds) and
VDM thermodynamic estimator (full circles)). 
The long dashed line indicates 
the theoretical high pressure limit $\rho=4 \rho_0$ of the fully dissociated non-interacting plasma.
In the experiments, a shock wave propagates through a sample of precompressed liquid deuterium 
characterized by its initial state, ($E_0$,~$V_0$,~$p_0$). 
Assuming an ideal shock front, the variables of the shocked material ($E$,~$V$,~$p$) satisfy 
the Hugoniot relation \cite{Ze66},
\begin{equation}
H = E-E_0+\frac{1}{2}(V-V_0)(p+p_0)=0 \quad.
\end{equation}
The initial conditions in the experiment were $T=19.6\,\rm{K}$ and 
$\rho=0.171\,\rm{g/cm^3}$. We set $V_0=39.1\,\rm{\AA^3}$ and $p_0 \approx 0$. 
We show two VDM curves based on the thermodynamic and direct estimators. For $E_0$, we use the 
corresponding value of the ground state of the isolated hydrogen molecule, $E_0^{th}=-0.955\,\rm{Ha}$ and
$E_0^{dir}=-1.124\,\rm{Ha}$.

We expect the difference of the two estimators to give a rough estimate of the accuracy 
of the VDM approach. At high temperature, the difference is relatively
 small and agreement with RPIMC simulations is reasonable.
Both VDM estimators  indicate that there is maximal 
compressibility around 1.5 Mbar. However, in this regime of high density
and relatively low temperature a more careful study seems unavoidable.
We suggest RPIMC simulations using the VDM nodal surface to restrict 
the paths.

\section{Conclusions}

    The VDM approach provides a way to systematically improve the many particle density matrix.
Already the simplest ansatz using one Gaussian to describe the single 
particle density matrices  gives a good description of hydrogen in the 
discussed range of temperature and density. The method 
includes the correct high temperature behavior and shows the expected formation of atoms 
and molecules. The thermodynamic variables are in reasonable agreement with RPIMC simulations.
  The presented Gaussian ansatz can be improved in several ways. One could use a sum of 
Gaussians, add underestimated correlation effects by including a Jastrow factor in the ansatz or
use a two-step path integral. Further one can use this essentially analytic density matrix
to furnish the nodal surface in RPIMC simulations, replacing the free
particle nodes by a density matrix that already includes the principle 
physical effects.
This level of accuracy seems to be required to determine a Hugoniot function 
that is very sensitive to the different level of approximations made by various theories.

\acknowledgements
The authors would like to thank David Ceperley for useful discussions.
This work was partially supported by the CSAR program and 
performed under the auspices of the U.S. Department of
Energy by Lawrence Livermore National Laboratory under contract
No. W-7405-Eng-48.
\appendix
\section{Gaussian Approximation Interaction Terms}

     The general equations for the variational parameters ${q}$ in a 
parameterized density matrix, from Eq.~\ref{matrix}, are
\beq
      \label{eqom}
\frac{1}{2}\frac{\partial H}{\partial \vec{q}}\; + \;\, 
    \stackrel{{\textstyle \leftrightarrow}}{\NN}\: \dot{\vec{q}}=  0
\eeq
where
\beq
H \equiv \int \rho \HH \rho\;d\RR = \int \rho \HH \rho_{I}d\RR 
                     \label{A2}
\eeq
and the norm matrix 
\beq
\NN_{ji} \equiv \int p_j \: p_i \: \rho^2 \, d\RR =\lim_{q'\rightarrow q}
\frac{\partial^2 N}{ \partial q_j \partial q'_i}
\eeq
with
\beq
N \equiv \int \rho(\RR,\vec{q}\,;\beta) \; \rho(\RR,\vec{q}\:'\,;\beta)\;d\RR\;\;.
\eeq
The subscript $I$ in Eq. \ref{A2}  indicates that only one $\rho$ needs
to be antisymmetric and the identity permutation can be used in the
other. (We are also dropping $1/N!$ prefactors which are the same for 
the  norm matrix
and thus cancel out.) This appendix contains the detailed formulae for 
these equations for a parameterized Gaussian density matrix applied to 
a Coulomb system.

      Repeating Eq. \ref{product} the parameterized  variational density
matrix is an anti-symmetrized product of one-particle density matrices,
\beq
\rho(\RR,\RRp,\beta) =\sum_{\cal{P}}\epsilon_{\cal{P}} 
                   \prod_{k} \rho_1(\rr_k,\rr'_{{\cal{P}}_k},\beta) = 
\sum_{\cal{P}}\epsilon_{\cal{P}} e^D \prod_{k} \:(\pi w_{{\cal{P}}_k})^{-3/2} \:
       \mbox{exp} \left\{ -\frac{1}{w_{{\cal{P}}_k}} (\rr_k-\mm_{{\cal{P}}_k})^2
            \right\}
\eeq
where the amplitude $D$ and the widths $w_{k}$ and means
${\bf m}_{k}$ are the variational parameters. The permutation sum is
 over all permutations
of identical particles (e.g. same spin electrons) and $\epsilon_{\cal{P}}=\pm 1$is the permutation signature.  The initial conditions 
are $w_k=0$, $\mm_k = \rr_k'$, and $D=0$.

      For this ansatz the generator of the norm matrix, 
\beq
N= \sum_{\cal{P}}\epsilon_{\cal{P}}\prod_{k}
[\pi (w_{k}+w_{{\cal P}_k}')]^{-3/2}
  \exp\left\{-(\mm_{k}-\mm_{{\cal P}_k} ')^{2}/(w_{k}+w_{{\cal P}_k}')\right\}
\exp( D+D')\;.
\eeq
For a periodic system the above equation also is summed over all periodic
simulation cell vectors, ${\bf L}$, with $\mm_{k}-\mm_{{\cal P}_k}'\rightarrow
\mm_{k}-\mm_{{\cal P}_k}'+{\bf L}$.
      Using this the components of the norm matrix are then:
\BEN
\NN_{DD} &=&\sum_{\cal{P}}\epsilon_{\cal{P}} \NP
\\
\NN_{{\bf m}_{i}D} &=&\sum_{\cal P}\epsilon_{\cal P} \left[
  {-2({\bf m}_{i}-{\bf m}_{{\cal P}_i})\over w_{i}+w_{{\cal P}_i}}
    \right] N_{{\cal P}}
\\
\NN_{w_{i}D} &=&\sum_{\cal P}\epsilon_{\cal P}
    \left({-1\over w_{i}+w_{{\cal P}_i}}\right) \left[
{3\over 2}-{(\mm_{i}-\mm_{{\cal P}_i})^{2} \over w_{i}+w_{{\cal P}_i}}
                                                \right]\NP
\\
\NN_{{\bf m}_{i}{\bf m}_{j}} &=&\sum_{\cal P}\epsilon_{\cal P} \left[
{2\delta_{j,{\cal P}_i}\stackrel{{\textstyle \leftrightarrow}}{I}
     \over w_{i}+w_{j}}+4
{({\bf m}_{i}-{\bf m}_{{\cal P}_i}) \over  (w_{i}+w_{{\cal P}_i})}
{({\bf m}_{j}-{\bf m}_{{\cal P}_{j}^{-1}}) \over  (w_{j}
                                      +w_{{\cal P}_j ^{-1}})}
\right] \NP
\\
\NN_{{\bf m}_{i}w_{j}} &=&\sum_{\cal{P}}\epsilon_{\cal{P}} \left[
{\delta_{j,{\cal P}_i}\over w_{i}+w_{j}}
+{1\over (w_{j}+w_{{\cal P}_{j}^{-1}})} \left({3 \over 2}
-{({\bf m}_{j}-{\bf m}_{{\cal P}_{j}^{-1}}) ^{2}\over (w_{j}+
                                 w_{{\cal P}_{j}^{-1}})} \right)\right]
\left[ {2({\bf m}_{i}-{\bf m}_{{\cal P}_i})\over w_{i}+w_{{\cal P}_i}}
\right] \NP
\\
\nonumber
\NN_{w_{i}w_{j}} &=&\sum_{\cal{P}}\epsilon_{\cal{P}} \left\{
{\delta_{j,{\cal P}_i}\over (w_{i}+w_{{\cal P}_{i}})^{2}}
\left[ {3 \over 2} - {2(\mm_{i}-\mm_{{\cal P}_{i}})^{2}\over
    w_{j}+ w_{{\cal P}_{j}}} \right]\right.
    +{1\over (w_{i}+w_{{\cal P}_i})(w_{j}+w_{{\cal P}_{j}^{-1}})}
\\ 
& &  \hspace*{.5in} \left.
\left[{3 \over 2}- {(\mm_{i}- \mm_{{\cal P}_i})^{2}\over
      w_{i}+w_{{\cal P}_i}}\right]
\left[{3 \over 2}- {(\mm_{j}- \mm_{{\cal P}_{j}^{-1}})^{2}\over 
                 w_{j}+w_{{\cal P}_{j}^{-1}}}
   \right] \right\} \NP
\\
\nonumber
\mbox{where}&&
\\
\NP&=& e^{2D}\prod_{j} \begin{array}{c}
   \underline{ \exp \left\{
            -\frac{(\mm_{j}-\mm_{{\cal P}_j} )^{2}}{(w_{j}+w_{{\cal P}_j})}
        \right\}} \\ (\pi (w_{j}+w_{{\cal P}_j}))^{3/2}
                         \end{array} = N_{{\cal P}^{-1}}\;.
\EEN
     The Hamiltonian for a periodic system of electrons and ions
\beq
    \HH =-\frac{1}{2}\sum_{i=1}^{N_{e}} {\bf \nabla}^{2}_{i} +\sum \sum_{i<j}
          \psi({\bf r}_{ij}) -\sum_{i}\sum_{I} Z_{I}\psi({\bf r}_{iI})
          +\sum_{i}U_{Mad} +U_{ions}
\eeq
      where the purely  ionic terms
\beq
    U_{ions} = \sum\sum_{I<I'} Z_{I}Z_{I'}\psi({\bf r}_{II'})
          +\sum_{I} Z_{I}^{2}U_{Mad} \;.
\eeq
 The Ewald potential, $\psi(\rr)$, which includes interactions
with periodic images and incorporates charge neutrality,
\beq
      \label{ewald}
\psi(\rr)=\sum_{{\bf L}}\begin{array}{c}\underline{\erfc(G
                   |\rr+{\bf L}|)}\\ |\rr+{\bf L}|\end{array}+
     \sum_{{\bf k}\neq 0}{4\pi\over\Omega k^{2}}\exp(-k^{2}/4G^{2})-
      {\pi\over G^{2}\Omega}=\sum_{{\bf k}\neq 0}{4\pi\over\Omega k^{2}}
       \exp(i{\bf k}\cdot \rr)
\eeq
where  $\Omega$ is the 
periodic cell volume and $G$ an arbitrary constant. The Madelung term in $\HH$
is the interaction energy of an electron with it's periodic images and
neutralizing background (e.g.
$U_{Mad}=-1.41865/L$ for a simple cubic simulation cell, 
the usual case). 
      To do the integrals we represent the Gaussians by their Fourier series
\beq
({2\over \pi w})^{3/2}\sum_{{\bf L}} e^{-{2\over w}(\rr-\mm-
   {\bf L})^{2}}=
\sum_{{\bf k}}{1\over \Omega}e^{-k^{2}w/8}e^{i{\bf k}\cdot (\rr-\mm)}
\eeq
and in the interaction terms use the Fourier representation for $\psi(\rr)$.
      This finally gives
\beq
H=\sum_{\cal P} \epsilon_{\cal P} \left\{\KP+\UP\right\} \NP
\eeq
with
\BEN
 \KP &=&\sum_{i}\left[ \frac{3}{w_{i}+w_{{\cal P}i}} - 
 2\frac{({\bf m}_{i}-{\bf m}_{{\cal P}i})^{2}}{(w_{i}+w_{{\cal P}i})^{2}}\right ]
\\
\UP &=& \sum\sum_{i<j} W(\tilde{\bf m}_{i}-\tilde{\bf m}_{j},
         \tilde{w}_{i}+ \tilde{w}_{j})-\sum_{i}\sum_{I}Z_{I} 
         W(\tilde{\bf m}_{i}-\RR_{I}, \tilde{w}_{i}) +\sum_{i} U_{Mad}+U_{ions}
\EEN
where $ \tilde{w}_{i}\equiv w_{i}w_{{\cal P}i}/(w_{i}+w_{{\cal P}i})$ and
$ \tilde{{\bf m}}_{i}\equiv ({\bf m}_{i}w_{{\cal P}i}+{\bf m}_{{\cal P}i}
    w_{i})/ (w_{i}+w_{{\cal P}i})\;$ .
The interaction integral 
\beq
       \label{int_int}
 W({\bf r},w)\equiv
\sum_{k\neq 0}\frac{4\pi}{\Omega k^{2}}
e^{-\frac{k^{2} w }{4}}
e^{i{\bf k}\cdot {\bf r}}
\eeq
 $W$ is symmetric in ${\bf r}$ when the periodic
cell has inversion symmetry.
Continuing, the left hand side of Eq. \ref{eqom} is
\BEN
H_{D} &\equiv &\frac{1}{2}\frac{\partial H}{\partial D}=H
\\
H_{w_{i}} &\equiv &\frac{1}{2}\frac{\partial H}{\partial w_{i}}=
\frac{1}{2}\sum_{\cal P} \epsilon_{\cal P}\left\{
(\frac{\partial \KP} {\partial w_{i}}+\frac{\partial \UP}{\partial w_{i}})\NP+
(\KP+\UP) \frac{\partial \NP}{\partial w_{i}}\right\}
\\
H_{{\bf m}_{i}} &\equiv &\frac{1}{2}\frac{\partial H}{\partial {\bf m}_{i}}=
\frac{1}{2}\sum_{\cal P} \epsilon_{\cal P}\left\{ 
(\frac{\partial \KP} {\partial {\bf m}_{i}}+
\frac{\partial \UP}{\partial {\bf m}_{i}})\NP+
(\KP+\UP) \frac{\partial \NP}{\partial {\bf m}_{i}}\right\}
\EEN
with
\BEN
\frac{\partial \NP}{\partial w_{i}}&=& \left[-\frac{3}{w_{i}+w_{{\cal P}i}} +
2\frac{({\bf m}_{i}-{\bf m}_{{\cal P}i})^{2}}{(w_{i}+w_{{\cal P}i})^{2}}\right]\NP
\\
   \frac{\partial N_{{\cal P}}}{\partial {\bf m}_{i}}&=& \left[
  -4\frac{({\bf m}_{i}-{\bf m}_{{\cal P}i})}{w_{i}+w_{{\cal P}i}}\right] \NP
\\
\frac{\partial \KP}{\partial w_{i}}&=& \left[-\frac{6}{(w_{i}+w_{{\cal P}i})^{2}}+
     8\frac{({\bf m}_{i}-{\bf m}_{{\cal P}i})^{2}}{(w_{i}+w_{{\cal P}i})^{3}}\right]
\\
\frac{\partial \KP}{\partial {\bf m}_{i}}&=& \left[
-8\frac{({\bf m}_{i}-{\bf m}_{{\cal P}i})}{(w_{i}+w_{{\cal P}i})^{2}}\right]\;.
\EEN
where we have used the fact that terms in ${\cal P}i$ and ${\cal P}^{-1}i$ give the
same contribution under the permutation sum and so combined them.
The derivatives of the interaction integral are,
\BEN
   \frac{\partial \UP}{\partial {\bf m}_{i}}&=&
   \frac{2 w_{{\cal P}i}}{w_{i}+w_{{\cal P}i}}  
\left[
     \sum_{j\neq i}{\bf W}^{[1]}(\tilde{m}_{i}-
      \tilde{m}_{j},\tilde{w}_{i}+\tilde{w}_{j})-\sum_{I}Z_{I}
     {\bf W}^{[1]}(\tilde{m}_{i}-{\bf R}_{I},\tilde{w}_{i}) 
\right]
\\
\nonumber
\frac{\partial \UP}{\partial w_{i}}&=&
   \frac{2 w_{{\cal P}i}}{(w_{i}+w_{{\cal P}i})^{2}}  \left[
w_{{\cal P}i}\left(
   \sum_{j\neq i} W^{[2]}(\tilde{m}_{i}- \tilde{m}_{j},
\tilde{w}_{i}+\tilde{w}_{j})-\sum_{I} Z_{I} W^{[2]}(\tilde{m}_{i}-
     {\bf R}_{I}, \tilde{w}_{i})  \right)
\right. 
\\ 
& &  \hspace*{0.15in} + \left.  
({\bf m}_{{\cal P}i}-{\bf m}_{i})\cdot\left( \sum_{j\neq i}{\bf W}^{[1]}
(\tilde{m}_{i}- \tilde{m}_{j},\tilde{w}_{i}+\tilde{w}_{j})-\sum_{I}Z_{I}
 {\bf W}^{[1]}(\tilde{m}_{i}-{\bf R}_{I},\tilde{w}_{i})\right) 
\right]
\EEN
where ${\bf W}^{[1]}$ and $W^{[2]}$ denote the derivatives of $W$ with the first
and second argument.
Comparing equation \ref{int_int} and Eq. \ref{ewald} the interaction
integral  may be written as
\beq
W({\bf r},w)=
\psi({\bf r})-\sum_{\bf L} \begin{array}{c}\underline{
\erfc\left[\frac{|{\bf r}+{\bf L}|} {\sqrt{w}}\right]}
 \\ |{\bf r} +{\bf L}| \end{array}
+\frac{\pi w }{\Omega}
\eeq
and its derivatives as:
\BEN
 {\bf W}^{[1]}({\bf r},w)&=& {\bf\nabla}\psi({\bf r})+
          \sum_{\bf L} \frac{{\bf r}+{\bf L}}{|{\bf r}+{\bf L}|^{3}}
\left( \erfc\left[\frac{|{\bf r}+{\bf L}|} {\sqrt{w}}\right]
+\frac{2 |{\bf r}+{\bf L}|}{\sqrt{\pi w}}\exp(-|{\bf r}+{\bf L}|^{2}/w)\right)
\\
 W^{[2]}({\bf r},w) &=&-\sum_{\bf L} 
       \frac{\exp(-|{\bf r}+{\bf L}|^{2}/w)}{w^{3/2}\sqrt{\pi}}  +\frac{\pi}{\Omega}
\EEN
For an isolated system (${\bf L}\rightarrow\infty$) and these would 
simplify to,
\BEN
  W({\bf r},w)&=&\frac{\erf\;[r/\sqrt{w}\;]}{ r} 
\\
  {\bf W}^{[1]}({\bf r},w)&=&-\frac{{\bf r}}{r^{3}}\left(
          \erf\;[r/\sqrt{w}\;]-\frac {2 r}{\sqrt{\pi w}}e^{-r^{2}/w}\right)
\\
  W^{[2]}({\bf r},w)&=&-\frac{1}{w \sqrt{\pi w}}e^{-r^{2}/w}
\EEN

   At $\beta=0$ the initial derivatives for the variational parameters reduce to
\BEN
      \dot{w}_{i}&=&2   \\
      \dot{\bf m}_{i}&=&0 \\
      \dot{D}&=&-U_{I}
\EEN

      For large numbers of electrons it is not possible to treat all permutations.
Here the approximation discussed in section VII is used where the kinetic pair exchange 
corrections given there are added to the identity permutation term derived here.

\bibliographystyle{unsrt}
\bibliography{bloch}
\end{document}